\documentclass{jpsj-suppl}
\usepackage{txfonts} 

\title{Recent Progress in Cosmology and Particle Astrophysics}

\author{Pisin \textsc{Chen}$^{1,2,3,4}$}

\inst{
$^1$ Department of Physics, National Taiwan University, Taipei, Taiwan 10617\\
$^2$ Graduate Institute of Astrophysics, National Taiwan University, Taipei, Taiwan 10617\\
$^3$ Leung Center for Cosmology and Particle Astrophysics, National Taiwan University, Taipei, Taiwan 10617\\
$^4$ Kavli Institute for Particle Astrophysics and Cosmology, SLAC National Accelerator Laboratory, Stanford University, Stanford, CA 94305, U.S.A.
}

\email{pisinchen@phys.ntu.edu.tw; chen@slac.stanford.edu}

\recdate{September 1, 2013}

\abst{Recent years have seen dramatic progress in cosmology and particle astrophysics. So much so that anyone who dares to offer an overview would certainly risk him- or herself for being incomplete and biased at best, and even incorrect due to the author's limited expertise. It is with this understanding and excuse that I risk myself in offering this review. After a brief summary of Planck mission's first results, I highlight some selected theoretical and experimental advancement in dark energy, dark matter, and cosmic neutrinos research. It is hoped that with a glance through these exciting development, one would be convinced that we are now a step closer to the ultimate understanding of our universe, while major breakthroughs are still required.}

\kword{cosmology, particle astrophysics, inflation, dark energy, dark matter, cosmic neutrinos}

\begin{document}
\maketitle

\section{Introduction}

Cosmology and particle astrophysics have enjoyed tremendous progress throughout the past two decades. Healthy interplays between observational/experimental data from numerous projects and theoretical concepts have greatly enhanced our fundamental understanding of the universe. Such advancement in turn triggers new questions to be further addressed. It is fare to say that we are currently living in an era of renaissance in cosmology and particle astrophysics. Under this light, it is clear that a brief overview talk cannot possibly cover all the major advancements nor can it serve justice to all the important works in this field. The situation is further worsened due to the author's limited expertise and partial knowledge in these burgeoning fields. At the risk of being partial and incomplete, we showcase the progress by a few selected topics with the hope that they would provide a good sampler to the much fuller tapestry of the advancement of these very exciting fields.

Modern cosmology began nearly 100 years ago with Einstein's finalizing his general theory of relativity (GR) and applying it to cosmology on the theory side and Hubble's discovery of cosmic expansion on the experimental side. Within a century, especially in the past 20 years, a {\it standard model of cosmology} has become mature, based on which we are now able to search for the remaining missing pieces. Yet certain aspects of this overall successful picture of the universe are puzzling enough that they may require new concepts that would go beyond this standard model. 

Our overview focuses on four topics: early universe, dark energy (DE), dark (DM) , and high-energy cosmic neutrinos. We will highlight the progress of each in the following sections. In the areas of early universe and the dark side of the universe, the recently released Planck data has provided refinement on key parameters in cosmology and tighter constraints on theoretical models. The nature of DE, however, remains illusive. We will review a few theoretical attempts of addressing this issue. Recent progress in DM search comes primarily from the experimental side in all three channels of DM detection. Results from space-based indirect search, however, appears to be inconsistent with that from underground direct search of DM. Some resolution is evidently required. Finally, there are also exciting developments from the cosmic neutrino front. It is hoped that with a glance through these exciting developments, one would be convinced that we are now a step closer to the ultimate understanding of our universe, while major breakthroughs are still required.




\section{Early Universe}

\subsection{Summary of Planck Results}
On the whole, Planck has not (yet) discovered any major discrepancy from the notions of the standard model of cosmology \cite{Planck-I}.With much better precisions than its predecessor, WMAP, it indicates instead semi-significant changes of the value of several key parameters as well as constraints on inflation. Here we review Planck's major results. 

\noindent$\bullet$ The universe is expanding slower than previously thought, with the Hubble parameter at present determined as $H_0=67.3\pm 1.2 {\rm km/s/Mpc}$ \cite{Planck-XVI}.

\noindent$\bullet$ The universe has slightly more dark matter and slightly less dark energy than previously estimated, with $\Omega_{DM}\simeq 26.8\%$ rather than $22.7\%$ and  $\Omega_{DE}\simeq 68.3\%$ rather than $72.8\%$ previously determined by WMAP \cite{Planck-XVI}.

\noindent$\bullet$ The simplest inflation models, i.e., $V(\phi)\propto \phi^n$, where $n>1$, are in more danger of being ruled out \cite{Planck-XXII}.

\noindent$\bullet$ The CMB anisotropy is found lower than expected at large scales \cite{Planck-XV}.

\noindent$\bullet$ There are likely 3 flavors of neutrino instead of 4 as previously suggested. The sum of their masses is very small (less than 0.18 eV). There is no evidence for sterile neutrino, though it is not conclusively excluded.

\section{Dark Energy}
\subsection{Dynamical Field vs. Cosmological Constant}
While the existence of a substance permeating in the universe by the name {\it dark energy} responsible for its accelerating expansion has been established, the nature of this substance is still undetermined. The cosmological constant (CC), initially introduced by Einstein, which is uniform and time-independent, is a natural candidate for dark energy. However it is also possible that the dark energy is provided by some dynamical fields. The density $\rho$ and pressure $p$ of dark energy can be described by the equation of state of ideal fluid,
\begin{equation}
p=w\rho,
\label{e1}
\end{equation} 
where $w$ is a dimensionless parameter that characterizes the nature of the substance. Cosmological constant is characterized by $w=-1$. Hubble expansion can be expressed in terms of the rate of change of the {\it scale factor} $a$: $H\equiv\dot{a}/a$. According to the Friedmann equation, the acceleration (or deceleration) of the universe, i. e., the rate of change of $H$, is governed by
\begin{equation}
\frac{\ddot{a}}{a}=-\frac{4\pi G}{3}(\rho+3p)+\frac{1}{3}{\Lambda}\,
\label{e2}
\end{equation}
where $\Lambda$ is the cosmological constant. We quickly see that the universe can undergo an accelerated expansion, i.e., $\ddot{a}/a>0$, either by having a positive $\Lambda$ or some dominant substance with $\rho+3p<0$. In terms of the equation of state parameter, this corresponds to $w<-1/3$. Evidently, CC, though simple and natural, is not the only possible solution to dark energy. One effective way to decipher the nature of dark energy is to trace the time, or equivalently scale factor, dependence of $w$. To this end one may parameterize $w$ as
\begin{equation}
w=w_0+w_a(1-a).
\label{e3}
\end{equation}
CC corresponds to $w_0=-1$ and $w_a=0$. 

The latest values from the Planck data give 
\begin{equation}
w_0=-1.04^{+0.72}_{-0.69},\quad\quad w_a<1.32, \quad\quad {\rm (95\%; Planck+WMAP+BAO)}.
\label{e4}
\end{equation}
We note that while Planck's result for dark energy equation of state does cover CC, other possible solutions based on dynamical fields and are therefore time-dependent, are not ruled out.

\subsubsection{Cosmic Acceleration}
As explained above, any dominant substance with $w<-1/3$ would accelerate the Hubble expansion. These can be subdivided into three categories:

\noindent$\bullet$  Quintessence: $\quad\quad\quad\quad\quad -1<w<-1/3.$

\noindent$\bullet$  Cosmological constant: $\qquad\quad\ \ w=-1,$

\noindent$\bullet$  Phantom energy:  $\quad\quad\quad\quad\quad\quad\ w<-1.$

\noindent The major difference between these three substances lies in the time-dependence of their energy densities. While the energy density of CC remains constant all the time, that of quintessence and phantom energy decreases and increases with time, respectively. This implies that if the dark energy is actually a phantom energy, then in the future, about 35 Billion years from now, the space-time of the universe will be ripped apart and the universe will end. The recent Planck result tends to lean slightly towards the phantom solution for dark energy, but this tendency is not yet statistically significant.

\subsection{Cosmological Constant: Problem and Solution}
So far CC remains a viable candidate for dark energy. It is simple and natural, readily incorporated in Einstein's general relativity. However, it also faces a long-standing {\it cosmological constant problem}. This CC problem stems from the tremendous discrepancy between the theoretical value associated with quantum vacuum energy and the value required to conform with observations. With a UV cutoff at the Planck scale ($\sim10^{19} {\rm GeV}$), the quantum vacuum energy density is $\sim 10^{112} {\rm eV}^4$, some 124 orders of magnitude larger than that of the {\it critical density} $\rho_c$ of the universe! So the CC problem before the discovery of the accelerated cosmic expansion in 1998 was mainly to find ways to put CC to zero. Since 1998, we recognize that the amount of dark energy (density) required to account for the acceleration should be $\rho_{DE}\sim (3/4)\rho_c$ at present. So the new challenge is why CC should be so small but nonzero. 
One obvious solution to the CC problem is that quantum vacuum energy should not gravitate. But why not? One possible answer is to develop a successful gravity theory in which cosmological constant is a constant of integration and is therefore disassociated with the vacuum energy. Here are couple attempts.

\subsubsection{Case 1: Unimodular Gravity}
This idea was first proposed by many authors in the early 1980s (for a review, see \cite{Weinberg:1989} and revived after 1998 \cite{Finkelstein:2001,Ellis:2013}. Its starting point is the same as Einstein-Hilbert action, 
\begin{equation}
S_{EH}=\frac{1}{8\pi G}\int dx^4 \sqrt{-g}R,
\label{e5}
\end{equation}
but with the constraint $\sqrt{-g}=1$; the conservation of energy-momentum is separately imposed. As a result, CC emerges as a constant of integration. In comparison, in Einstein's GR the CC is put in by hand without any constraint a priori. Although one still needs to fix this constant of integration, it nevertheless is disengaged from the quantum vacuum energy. One recent development along this line of approach is the demonstration by G. Ellis that unimodular gravity can accommodate the scenario of inflation \cite{Ellis:2013}. 

Although in this approach CC needs not be associated with quantum vacuum energy, the challenge remains as to why the integration constant must be so small. That is, the issue of fine-tuning remains.

\subsubsection{Case 2: Higher Order Gravity}
Here the idea is to replace the Einstein-Hilbert action, which is linear in the curvature tensor $R$, with quadratic contraction of Riemann tensors. This idea is not new. Stephenson, Kilmister and Newman proposed a gravity theory analogous to the gauge theory \cite{Steph:1958,Kilmister:1961} around 1960, where the action is composed of quadratic curvature tensor terms without the linear one. The action gives not only a higher derivative field equation that is consistent with Einstein's GR but also additional constraint equations.
Yang put forward another gauge theory of gravity in 1974 that satisfies the $GL(n)$ symmetry group \cite{Yang:1974}
and succeeded in deriving the same higher derivative field equation without the additional constraint equations.
Following the convention, we shall call the higher derivative field equation of gravity the SKY equation. The SKY equation reproduces all solutions of the vacuum Einstein equation.
Later Camenzind proposed a matter current term for the SKY field equation \cite{Camenzind:1975}, although it was not deduced from an action. This task was fulfilled by Cook in 2008 \cite{Cook:2008}. We shall call the complete theory that includes both pure space and matter contributions the SKYC gravity. 

Like the unimodular gravity, higher order gravity was revived in recent years as an attempt to address the dark energy problem \cite{Cook:2008,Chen:2010}. In addition, the affine connection instead of the metric serves as the theory's dynamical variable. It has been shown \cite{ChenIzumiTung:2013} that this theory yields two constants of integration, where one acts as CC and the other plays the role of {\it dark radiation}, which is an extra bonus. 

There are also problems with higher order gravity. First, a consistent matter action is still lacking; the matter current proposed by Camenzind \cite{Camenzind:1975} is not general covariant \cite{ChenIzumiTung:2013}. Second, as typical of higher order field theories, there exist ghost modes in this gravity theory. It has been argued, however, that such ghost modes can be ameliorated \cite{Salam:1978,Stelle:1978,Antoniadis:1986,Chen:2010}. 

\subsection{Smallness of Dark Energy}
As commented above, even if one manages to disengage CC from quantum vacuum energy, indeed any form of energy, the issue of fine-tuning remains. Why is CC, if it is indeed responsible for dark energy, so much smaller than all the fundamental gauge interaction scales? Here we review two approaches that attempt to address this question. 

\subsubsection{Due to Holographic Principle}
This approach does not try to disengage CC from quantum vacuum energy, but instead invokes the holographic principle to suppress the quantum zero-point energy. It was first suggested by Cohen, Kaplan and Nelson \cite{Cohen:1999} and Horava and Minic \cite{Horava:2000} in the aftermath of the dramatic discovery of the accelerating expansion of the universe. It was further investigated by Thomas \cite{Thomas:2002} and Li \cite{LiMiao:2004,LiMiao:2013}. 

It is argued that the quantum zero-point energy dominated by short-distance cutoff inside a region $L$ must be bounded by the mass of a black hole at the same size. That is,
\begin{equation}
L^3\rho_{\Lambda}\leq LM_p^2, \quad\quad\quad (M_p^2\equiv \frac{1}{8\pi G}),
\label{e6}
\end{equation}
where $\rho_{\Lambda}$ is the energy density associated with the zero-point energy and in this idea, the CC. Identifying $L$ with the present Hubble radius, i.e., $L=1/H_0\sim 10^{28} {\rm cm}$, one readily finds that $\rho_{\Lambda}\sim 10^{-10} {\rm eV}^4$, just about right for CC. 

Since Hubble radius is time-dependent, it remains a wonder why our present epoch is so special as to determine the value of CC for all time. This, in a sense, is therefore a variant of the concordance problem.  

\subsubsection{Due to Gauge Hierarchy}
Another attempt to address the DE smallness problem is through the double suppression of {\it gauge hierarchy} \cite{Murayama,Chen:2009a}. It has long been recognized that among the four fundamental interactions there exists a huge gap between the three gauge interaction scale, $M_{SM}$, presumably at $\sim 1 {\rm TeV}$, and that of quantum gravity at Planck scale $M_p$, by about 16 orders of magnitude: $M_{SM}/M_p\sim 10^{-16}$. These authors noticed that the DE scale happens to be 
\begin{equation}
\rho_{DE}\sim \Big(\frac{M_{SM}}{M_p}\Big)^2 M_p \equiv \alpha_G^2 M_p.
\label{e7}
\end{equation}

Taking this as a guiding principle, a model based on Randall-Sundrum brane world (4+1)D geometry, RS1, was introduced \cite{Shao:2010}. It was demonstrated that a naturally small quantum correction, or the Casimir energy, to the cosmological constant can arise from a massive bulk fermion field in the Randall-Sundrum model that satisfies exactly Eq.(\ref{e7}) and thus provides the correct value of DE on the TeV brane.

\section{Dark Matter}
\begin{figure}[tbh]

\begin{center}
\includegraphics[width=0.5\textwidth]{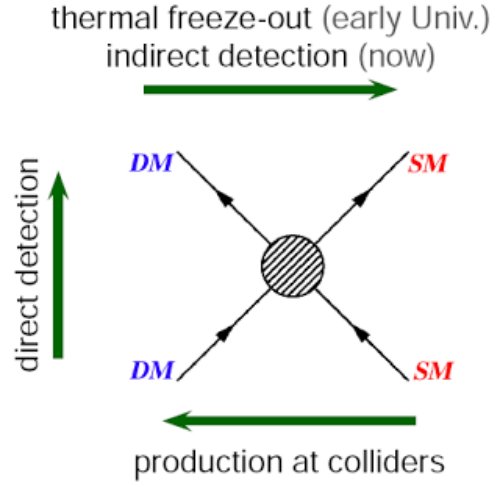}
\caption{Three possible channels to detect dark matter: direct detection, indirect detection, and production at colliders.}
\label{f1}
\end{center}
\end{figure}

Experimentally speaking, there are at least three ways to decipher the mystery of dark matter. These are direct detection, indirect detection and production at colliders. Direct detections are commonly carried out in underground laboratories, which rely on signals induced by the interaction of DM with ordinary matter. Indirect detections, often carried out in outer space, searches for relic signals due to DM-DM annihilation during thermal freeze-out in early universe. Collider search, such as that at LHC, looks for the production of DM from the interactions of standard model particles. Figure \ref{f1} explains the inter-relationship between these three different approaches.

\subsection{Direct Detection}
\begin{figure}[tbh]
\begin{center}
\includegraphics[width=0.7\textwidth]{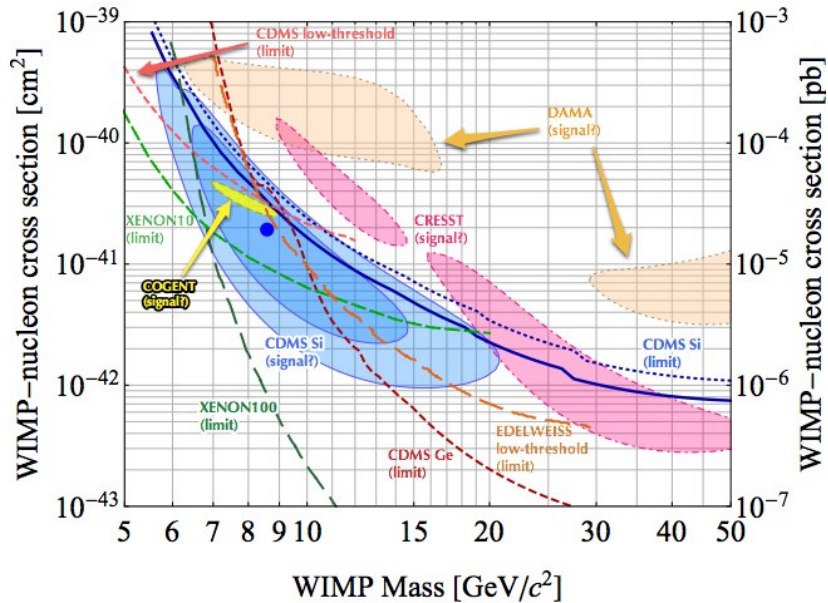}
\caption{Dark matter cross section as a function of mass (figure taken from \cite{CDMS}). Results from other experiments such as DAMA, CRESST,  CoGeNT, and XENON are also plotted for comparison.}
\label{f2}
\end{center}
\end{figure}

There have been exciting news in this front of search for DM. Earlier, DAMA/LIBRA's finding \cite{DAMA/LIBRA} of annual modulation hints on the existence of DM with a mass much lighter than what commonly assumed for WMIPs based on SUSY neutralino scenarios. Several experiments, namely CoGeNT \cite{CoGeNT} and CRESST \cite{CRESST}, have independently reported on the evidence of light-mass DM at 7-10 GeV. More recently, CDMS collaboration has reported the observation of three nuclear recoil-like events with a best fit corresponding to a DM mass of $m_{\rm DM}=8.6{\rm GeV}$ and cross section $\sigma{SI}=1.9\times 10^{-41}{\rm cm}^2$, consistent with CoGeNT's finding \cite{CDMS}. Constraints from XENON100 experiment \cite{XENON100}, on the other hand, appear to be incompatible with such hypothesis. Figure \ref{f2} compiled by CDMS-II \cite{CDMS}, shows the constraints reported by these experiments. It has been argued \cite{Hooper}, however, that with lower, but not implausible, values for the relative scintillation efficiency of liquid xenon ($L_e$), and the suppression of the scintillation signal in liquid xenon at XENON100's electric field ($S_{nr}$), the two events found by XENON could consistently arise from dark matter particles with a mass and cross section in the range favored by CoGeNT and CDMS.

\subsection{Indirect Detection}
Recent years have also seen exciting advancements in the indirect detections of dark matter, spearheaded by PAMELA's discovery of excess positron-to-electron flux ratio above $\sim 10{\rm GeV}$ \cite{PAMELA}. This finding was confirmed by Fermi LAT \cite{Fermi1, Fermi2} and more recently AMS-2 onboard International Space Station \cite{AMS}. In addition, the hardening of the individual electron/positron spectrum was also found. These may be attributed to DM-DM annihilation: $\chi\chi \to e^+e^-$,  or decay: $\chi \to e^+e^-$, but it may also be contributed by astrophysical objects such as pulsars. Combined analysis of PAMELA, Fermi LAT, and AMS-2 data with various assumptions on the background puts the DM mass at around TeV level \cite{Feng:2013}. This is consistent with an earlier analysis based on PAMELA, Fermi and ATIC, which suggested a DM mass at 700+ GeV \cite{Cholis:2009}. Such conclusion, if true, would be in sharp contrast with the low DM mass determined by several direct detection experiments. With this in mind, there has been the suggestion that the observed positron access can be well explained by a local, middle-aged supernova remnant and has nothing to do with DM \cite{Erlykin:2013}.
 
\subsection{Collider Search}
The lightest stable particle in supersymmetry (SUSY) has been a popular candidate for weakly interacting massive particle (WIMP) dark matter. However SUSY has not been found in colliders including LHC \cite{ATLAS,CMS}. This plus the discovery of Higgs at 126GeV have imposed constraints on the Constrained Minimal Supersymmetric Standard Model (CMSSM) \cite{Bechtle:2012}, and the standard neutralino abundance tends to be far above WMAP-measured values \cite{Baer:2012}. In light of the recent evidence of direct detection of DM as light as 6-10GeV described above, results from LHC also raised the concern whether SUSY neutralino remains a viable candidate for DM with such low mass. It has been suggested recently that the Next-to-MSSM (NMSSM) may be able to cover this mass range for neutralino under some maneuvering of the model \cite{Kozaczuk:2013}. Without invoking the notion of SUSY but assuming the CDMS-II events are indeed DM, Cheung et al. have suggested possible collider signatures to be searched at LHC \cite{Cheung:2013}.

\section{Cosmic Neutrinos}
After reviewing the two major components of our present universe, dark energy and dark matter, which together comprise $\sim 95\%$ of the total substance, we now turn to the remaining 5\%, the ordinary matter, whose fundamental properties are by and large understood under the standard model of particle physics. This, however, does not mean that there is no more mystery left, in the cosmic context. Among all elementary particles, neutrinos are arguably the least understood. It was not until recently that we recognized that neutrinos oscillate between their 3 flavors and that they are actually not massless. It seems therefore prudent to reexamine the evolution of the pre-recombination early universe on the possible impacts due to these salient properties of neutrinos. There are also potential impacts on late universe, not on large-scale cosmological effect but on neutrino detections that may shed lights on fundamental properties of neutrinos such as neutrino decay. 


\subsection{Decay or Not Decay?}
The origin of ultra-high energy cosmic rays (UHECR) has been widely regarded as one of the major questions in the frontiers of particle astrophysics. Gamma ray bursts (GRB), the most violent explosions in the universe second only to the Big Bang, have been a popular candidate site for UHECR productions. The recent IceCube report on the non-observation of GRB induced neutrinos \cite{IceCube} therefore attracts wide attention. This dilemma requires a resolution: either the assumption of GRB as UHECR accelerator is to be abandoned or the expected GRB induced neutrino yield was wrong. It has been pointed out by Zhuo Li that IceCube has overestimated the neutrino flux at GRB site by a factor of $\sim 5$ \cite{LiZhuo}. In addition to the issue of neutrino production at source, the neutrino oscillation and the possible neutrino decay during their flight from GRB to Earth should further reduce the detectability of IceCube, which is most sensitive to the muon-neutrino flavor as far as point-source identification is concerned. 

In addition to oscillations, another major discovery about neutrinos in the last two decades is that neutrinos have mass. This, together with additional assumption of the decay process based on notions beyond the standard model of particle physics, leads to the prediction of neutrino decay, from the higher mass eigenstates to the mass ground state. One can envision at least two possible mass hierarchies: the normal hierarchy, i.e., $m_3\gg m_2> m_1$, and the inverted hierarchy, i.e., $m_2> m_1 \gg m_3$. It has been shown that under normal hierarchy the eventual flavor ratio on Earth, normalized to number of neutrinos produced at GRB site, is \cite{Beacom:2003,Maltoni:2008}
\begin{equation}
{\rm \quad Normal\ Hierarchy:}  \quad\quad\quad f^E_{e}:f^E_{\mu}:f^E_{\tau}=2/3:1/8:5/24,
\end{equation}
or 
\begin{equation}
{\rm Inverted\ Hierarchy:}  \quad\quad\quad f^E_{e}:f^E_{\mu}:f^E_{\tau}=0:2/5:3/5.
\end{equation}
So for the IceCube search for GRB emission of neutrinos, which is most sensitive to $\nu_{\mu}$, its detector sensitivity would be further reduced from 1/3 (based on pure oscillation) to 1/8 of the total neutrino flux for the case of normal hierarchy, which would be quite a sizable impact on the IceCube detector sensitivity, assuming that IceCube solely relies on muon-neutrinos in its investigation of the cosmic accelerator question. 

There are, however, new types of neutrino observatories based on Askaryan effect that are less sensitive to neutrino flavors and would therefore provide almost one order of magnitude improvement in neutrino detection sensitivity. As a trade-off, their angular resolution, on the other hand, will not be as good as that with muon-neutrino at IceCube. If such detector can in addition distinguish different flavors, then the measured flavor ratio would provide crucial information not only on the nature of UHECR production at source, but also on the nature of neutrino decay \cite{Chen:2012}. 

\subsection{Ultra-High Energy Cosmic Neutrino Detection}
We now turn to the detection side of ultra-high energy cosmic neutrinos. We review three experiments: ANITA \cite{ANITA-1,ANITA-2}, IceCube \cite{IceCube}, and ARA \cite{Chen:2009,ARA:2011}. 

ANITA is a balloon-borne radio wave antenna detector searching for neutrino-induced Askaryan signals emitting from Antarctic ice. It has so far carried out two missions in the 2006-2007 and 2008-2009 Austral summer, respectively. One major discovery of ANITA-1 was the observatory of ultra-high energy cosmic rays (UHECR) instead of neutrinos. Within about one month of flight, it has detected 16 such events with energy above $10^{19}$eV \cite{ANITA-1}. ANITA-2 has identified one candidate UHE neutrino event and determined the latest upper bound of UHE cosmic neutrino flux \cite{ANITA-2}. 

Aside from the non-observation of neutrinos from GRB discussed above, IceCube has recently published the observation of three neutrino events at PeV energy \cite{IceCube:2013}. This is very exciting. We expect more exciting news from IceCube. 

To go beyond ANITA and IceCube, a new international ARA (Askaryan Radio Array) Collaboration was formed with the ambition to bring the very effective detection mechanism based on Askaryan effect back on ground at South Pole, to cover 200 ${\rm km}^2$ area with 37 radio antenna stations. So far 3 stations have been deployed and more are under development. If ARA can indeed be built according to plan, it will complement IceCube by extending the UHE neutrino detection to higher energy range above $10^{17}{\rm eV}$. As all three neutrino flavors can trigger Askaryan signals, albeit with different characteristics, ARA can in principle distinguish different flavors given sufficient statistics\cite{Wang}. This would then provide an exciting prospect of shedding some light on the very fundamental issue of neutrino decay. 

\section{Summary}
Though extremely brief in content and selective in topics, this overview hopefully has provided a good sampler on recent progress in cosmology and particle astrophysics. By now we have a highly precise CMB measurement through the Planck mission. Although nothing dramatically deviates from our previous understanding, we are now certain that dark energy and dark matter must exist. However their nature are not yet fully unveiled. Even in the ordinary matter sector, the nature of neutrinos and their impact on cosmology are not yet fully understood. Historical breakthroughs in our understanding of the universe are destined to occur in years to come.

\end{document}